\documentclass[prb,twocolumn,amssymb,amsmath,aps,showpacs,floatfix,makecell]{revtex4-2}
\usepackage[colorlinks=true,linkcolor=blue,citecolor=blue,urlcolor=blue]{hyperref}
\usepackage{graphicx,bm}

\begin{document}

\title{Ballistic flow of two-dimensional electrons in a magnetic field}

\date{\today}

\author{A.~N.~Afanasiev}
\email{afanasiev.an@mail.ru}
\author{P.~S.~Alekseev}
\author{A.~A.~Greshnov}
\author{M.~A.~Semina} 
\affiliation{Ioffe Institute, St.~Petersburg 194021, Russia}

\begin{abstract}
In conductors with a very small density of defects, electrons at low temperatures collide predominantly with the edges of a sample. Therefore, the ballistic regime of charge and heat transport  is realized. The application of a perpendicular magnetic field substantially  modifies the character of ballistic transport. For the case of two-dimensional (2D) electrons in  the magnetic fields corresponding to the diameter of the cyclotron trajectories smaller than the sample width a hydrodynamic transport regime is formed. In the latter regime, the flow is mainly controlled by rare electron-electron collisions, which determine the viscosity effect. In this work, we study the ballistic flow of 2D electrons in long samples in magnetic fields up to the critical field of the transition to the hydrodynamic regime. From the solution of the kinetic equation, we obtain analytical formulas for the profiles of the current density and the Hall electric field far and near the ballistic-hydrodynamic transition as well as for the longitudinal and the Hall resistances in these ranges. Our theoretical results, apparently, describe the observed longitudinal resistance of pure graphene samples in the diapason of magnetic fields below the ballistic-hydrodynamic transition.
\end{abstract}

\maketitle


\section{Introduction}
In rather small samples of pure two- and three-dimensional conductors, electrons at very low temperatures most often collide with the edges of a sample, and therefore, their transport is ballistic. As the temperature increases, electron--electron collisions can lead to the formation of a viscous electron fluid and the implementation of hydrodynamic transport. Although the theory of a viscous electron fluid has been intensively developed for a long time~\cite{Gurzhi, Spivak, Andreev}, undoubted experimental evidence for the formation of such a fluid was obtained only recently in high-quality graphene, Weyl semimetals, and GaAs quantum wells~\cite{Weyl_sem_1, graphene, Levitov_et_al, Gusev_2, graphene_2, graphene_3, rrecentnest, rrecentnest2, Weyl_sem_2, exps_neg_3, exps_neg_1, exps_neg_2, exps_neg_4, Gusev_1, Gusev_2_Hall, je_visc, Al_Dm, exp_GaAs_ac_1, exp_GaAs_ac_2, exp_GaAs_ac_3, vis_res, Alekseev_Alekseeva}. The formation of a hydrodynamic flow in these experiments manifested in specific dependence of the average resistance of a sample on its width~\cite{Weyl_sem_1}, nonlocal negative resistance~\cite{graphene, Levitov_et_al, Gusev_2}, giant negative magnetoresistance~\cite{Weyl_sem_2, exps_neg_3, exps_neg_1, exps_neg_2, exps_neg_4, Gusev_1, Gusev_2_Hall, je_visc, Al_Dm}, and magnetic resonance at the double cyclotron frequency~\cite{exp_GaAs_ac_1, exp_GaAs_ac_2, exp_GaAs_ac_3, vis_res, Alekseev_Alekseeva}. The experimental implementation of hydrodynamic transport has led to the development of its theory in new directions (see, for example,~\cite{Lucas, eta_xy, we_4,recentest, we_5_1, we_5_2, Khoo_Villadiego, Kiselev_Schmalian_1___bound}).

In recent works~\cite{rrecentnest,rrecentnest2}, the spatial distribution of the current density and Hall electric field in a flow of two-dimensional (2D) electrons in graphene strips was measured. In~\cite{rrecentnest}, the observed evolution of the
Hall-field profile curvature served as evidence for a transition between the ballistic and hydrodynamic transport regimes. In~\cite{rrecentnest2} the current density of hydrodynamic and Ohmic flows in a narrow strip was determined by measuring the distribution of a local magnetic field induced by the current in the strip.

In~\cite{we_6,we_6_2}, the flow of interacting 2D electrons in a narrow ballistic sample was theoretically investigated in the limit of weak magnetic fields using the analytical solution of a simplified kinetic equation. It was shown that, at low temperatures, the Hall electric field in almost the entire sample, except the very edge vicinities, is half of its usual value in macroscopic ohmic samples (the exception of the edge vicinities was established in~\cite{a}). In addition, it was demonstrated in~\cite{we_6,we_6_2} that interparticle collisions, first, control the maximum ballistic trajectory length, which determines the ballistic current and the Hall electric field and, second, lead to small hydrodynamic corrections, which are precursors of the formation of a viscous flow.

In~\cite{Scaffidi}, the transition between the ballistic and hydrodynamic transport regimes for the Poiseuille flow of 2D electrons in a perpendicular magnetic field was theoretically investigated using numerical solution of the kinetic equation. In the magnetic field $B=B_c$, in which the diameter of the electron cyclotron orbit  $2R_c$ becomes equal to the sample width $W$ and some electrons start to make a complete revolution without scattering at the edges, the longitudinal and Hall resistances as functions of the magnetic field experience a jump. In~\cite{pohozaja_statja}, the Poiseuille flow of 2D electrons in a magnetic field was studied in more detail; specifically, the Hall-field distributions over the sample cross section were numerically calculated for $B<B_c$ and $B>B_c$ and an analytical solution of the kinetic equation in the ballistic region $B<B_c$ was constructed. This allowed the authors to explain the evolution of the Hall-field profile curvature upon a variation in the magnetic field, which was observed experimentally in~\cite{rrecentnest}. In~\cite{a}, an analytical theory for the transition between the ballistic and hydrodynamic transport regimes in a magnetic field at $|B-B_c|\ll B_c$ was developed. It was shown that, for a 2D electron flow in high-quality samples, this is an abrupt phase transition. The electron dynamics in the vicinity of the transition $|B-B_c|\ll B_c$ was revealed and a mean field model for quantitative description of the transition was constructed.

\begin{figure}[t!]
	\includegraphics[width=0.95\linewidth]{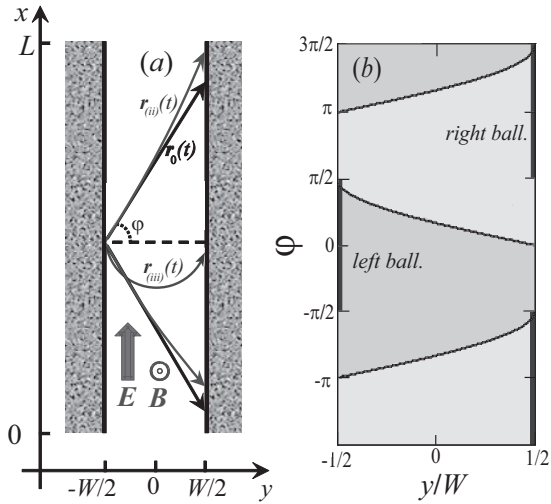}
		\caption{(a) Long ballistic sample in a magnetic field. The straight lines $\mathbf{r}_0(t)$ show the electron trajectories in zero magnetic field, $\mathbf{r}_{(ii)}(t)$ are the trajectories in weak fields corresponding to the magnetoresistance anomalies (second ballistic subregime), and $\mathbf{r}_{(iii)}(t)$ are the trajectories in relatively strong magnetic fields, which approach the phase transition between the ballistic and hydrodynamic regimes (the third ballistic subregime) from the lower side. (b) Areas in the coordinate plane $(y,\varphi)$, where mainly electrons reflected from the left and right sample edges move. Vertical thick lines show the position of the electrons just reflected from the left and right edges. The lines at the boundaries of the regions correspond to the trajectories touching the sample edges at angles of $ \varphi =\pm \pi/2 $.}
	\label{Fig1}
\end{figure}

The aim of this study is to theoretically investigate the ballistic transport of interacting 2D electrons in high-quality long samples in magnetic fields $B$ lower than the field $B_c$ of the transition to the hydrodynamic regime. Using the general analytical solution of the kinetic equation in the ballistic regime after~\cite{pohozaja_statja,a}, we show that, in the field range of $ B_1 < B < B_2 $ (more precisely, $B \gg B_1 $ and $  B_c-B \gg   B_c - B_2  $), the electron flow at low temperature is mainly determined by electron scattering at the sample edges and the cyclotron effect of a magnetic field. Weak electron--electron collisions with the intensity $\gamma$ determine the $B_1 $ and $B_2$ values ($B_1 \sim \gamma^2 $ and $  B_c - B_2\sim \gamma $) and only lead to minor corrections to all the characteristics of the flow in this range.

We calculate the current density and Hall field, as well as the longitudinal and Hall resistances  $\varrho_{xx}(B)$ and $\varrho_{xy}(B)$. With an increase in the magnetic field from $B_1$ to $B_2$,  the current density profile $j(y)$ evolves from almost flat to a deformed semicircle. The obtained Hall-field profiles $E_H(y)$, both in low fields ($B_1 \ll B \ll B_c$) and near the transition ($ B_c - B_2 \ll B_c-B \ll B_c $) are nonplanar and diverge at the sample edges: $ E_H(y) \neq  const $ and $ E_H (y)  \to \infty $ at $ y \to \pm W/2 $. The Hall-field amplitude becomes independent of magnetic field B in the region of moderately low fields ($B_c-B_2 \ll B_c-B \ll B_c $); therefore, the $ \varrho_{xx}(B)$  value in this region is plateau-like. This $\varrho_{xy}(B)$ behavior is apparently a form of the ballistic anomaly of magnetoresistance, which was previously observed experimentally and obtained by numerical simulation for 2D samples with four contacts~\cite{Beenakker_Houten_obz}. The magnetic-field dependencies of the resistances~ $\varrho_{xx}(B)$ and $\varrho_{xy}(B)$ and their derivatives with respect to $B$ over the entire range of $B_1 < B < B_2$ exhibit a nontrivial non-monotonic behavior and agree with the results of numerical simulation~\cite{Scaffidi}. The theoretical result for $\varrho_{xx}(B)$ apparently corresponds to the dependencies of the resistance of high-quality graphene samples and GaAs quantum wells experimentally observed in~\cite{rrecentnest,rrecentnest2,Gusev_2_Hall}.


\section{Model}

We consider a flow of 2D electrons in a high-quality long sample with the length $L$ and rough edges at low temperature. The scattering of electrons at the edges is assumed to be diffusive: the momentum of reflected electrons is isotropically distributed regardless of the momentum direction of incident electrons. In the bulk of the sample, electrons are rarely scattered by each other and/or weak disorder (Fig.~\ref{Fig1}a).

Our approach allows one to consider systems where two mechanisms of bulk scattering is present: (i) momentum-conserving electron--electron collisions and (ii) electron scattering by disorder, which leads to weak momentum relaxation. We assume the rate of any scattering in the bulk of the sample to be low: $W  \ll l$, where $l=v_F /\gamma$ is the electron mean free path with respect to all scattering mechanisms in the sample volume, $\gamma$ is the total scattering rate, and $v_F$ is the electron velocity at the Fermi level.

In weak magnetic fields ($B<B_c$), when the diameter of the cyclotron orbit is larger than the sample width ($ 2R_c >W$), each electron is mainly scattered at the sample edges. Consequently, in the main order in $\gamma$ in such magnetic fields, the electron scattering regime is ballistic. In this case, scattering in the bulk can limit the time spent by electrons on ballistic trajectories~\cite{we_6,we_6_2} and slightly modify~\cite{a} the purely ballistic flow (when electron--electron collisions are neglected).

In strong magnetic fields ($B>B_c$) corresponding to $ W > 2R_c$, electrons are divided into two groups with qualitatively different dynamics: ''edge'' electrons, which mainly move along the ''jumping'' trajectories and only scatter at one of the sample edges and ''central'' electrons with trajectories which do not touch the edges~\cite{Scaffidi,a}. Near the transition field ($B-B_c \ll B_c $), the centers of the ''central'' electron trajectories lie in a narrow range of coordinates at the sample center, $|y| \ll W$; therefore, their fraction is much smaller than the fraction of ''edge'' electrons, so they are scattered mainly at ''edge'' electrons and/or the bulk disorder. The occurrence of the ''central'' electrons denotes the initial stage of the formation of the bulk (2D) electrons phase responsible for hydrodynamic/Ohmic transport at $W \gg R_c$~\cite{a}.

We are searching for the linear response of 2D electrons to a uniform electric field  $ \mathbf{E}_0||x $ in an external magnetic field $\mathbf{B}$ perpendicular to the sample plane (Fig.~\ref{Fig1}a). The corresponding 2D electron distribution function acquires a nonequilibrium part $\delta f(y,\mathbf{p}) = -f_F'(\varepsilon) f(y, \varphi,\varepsilon) \sim E_0$, where $f_F (\varepsilon )$ is the Fermi distribution function, $\varepsilon$ is the electron energy, $\varphi$ is the angle between the electron velocity $\mathbf{v} = v(\varepsilon) [\, \sin \varphi , \cos \varphi \, ]$ and the normal to the left sample edge, $\mathbf{p}=m \mathbf{v} $  is the electron momentum, and $m$ is the electron mass (Fig.~\ref{Fig1}a). There is no dependence of $\delta f$ on coordinate $x$ along the
sample, since $L \gg W$. In addition, we ignore the dependence of the electron velocity $v(\varepsilon)=|{\bf v}(\varepsilon)|$  and the nonequilibrium part of the distribution function  $\delta f(y,\mathbf{p})$ on the electron energy. Such a simplification is justified for a degenerate electron distribution. Below, we use units in which the characteristic electron velocity $v(\varepsilon) \equiv v_F$ and the elementary charge $e$ are set to unity. In the selected units, the coordinate, time, and reverse field $1/E_0$ have the same dimensionality.

The kinetic equation for the nonequilibrium distribution function $f(y ,\varphi)  $ takes the form
\begin{equation}
\label{kin_eq}
\cos\varphi \, \frac{\partial f}{ \partial y }
 - \sin \varphi \, E_0 - \cos \varphi \, E_H -
  \omega_c \, \frac{\partial f}{ \partial \varphi } =
 \mathrm{St} [f]
 \:,
\end{equation}
where  $\omega_c =eB /mc $  is the cyclotron frequency, $E_H =E_H(y) $ is the Hall electric field induced by the redistribution of electrons in a magnetic field, and the collision integral $\mathrm{St}[f] $ describes both the momentum-conserving electron--electron collisions and scattering at bulk disorder, which leads to momentum relaxation.

In this work, we use a simplified form of the electron--electron and disorder collision integrals
\begin{equation}
 \label{St_N__St_U}
 \begin{array}{c}
  \displaystyle
\mathrm{St}  [f]  = - \gamma \, f +
 \gamma_{ee} \hat{P} [f] + \gamma_{d} \hat{P}_0 [f]
   \: ,
 \end{array}
\end{equation}
which allows us to obtain an asymptotically accurate (with respect to $\gamma W \ll 1 $) analytical solution of the kinetic equation at $B<B_c$. Here $\gamma_{ee}$ and $\gamma_{d}$ are the rates of electron--electron scattering and scattering at disorder, $\gamma = \gamma_{ee} + \gamma_{d} $ is the total scattering rate, and $\hat{P}$ and $\hat{P}_0 $ are the projection operators for the $f(\varphi)$ functions onto the subspaces $\{1,e^{\pm i \varphi} \}$ and $\{ 1 \}$, respectively. Such a collision integral preserves the distribution function perturbations corresponding to the nonequilibrium density. Collision integral~\eqref{St_N__St_U} at $\gamma_d = 0 $ also describes momentum conservation in interparticle collisions. This form of $ \mathrm{St}[f] $ was used in~\cite{we_6, we_6_2} to study the ballistic transport of interacting 2D electrons at $B \to 0$ and in~\cite{a} to construct the mean field theory of the phase transition between ballistic and hydrodynamic transport regimes near~$B=B_c$.

We assume that the longitudinal edges of the sample are completely rough. Thus, the electron scattering at the edges is diffusive and the boundary conditions for the distribution function have the form~\cite{Beenakker_Houten_obz,we_6_2,a} (see also Figs.~\ref{Fig1}a,b): $f(- W/2, \varphi)=c_l$ at angles within $- \pi/2 < \varphi < \pi / 2$ and $ f( W / 2, \varphi) = c_r $ at $   \pi/2 < \varphi < 3 \pi /2 $, where the quantities $c_{l} = c_{l} [f] $ and $c_{r} = c_{r} [f]$ are proportional to the $y$-components of the incident electron flow on the left $(y= - W/2)$ and right $(y=  W/2)$ sample edges:
\begin{equation}
  \label{c_12}
  \begin{array}{c}
  \displaystyle
  c_l =  - \frac{1}{2} \int _{\pi/2} ^{3\pi/2}
   d \varphi' \: \cos\varphi' \, f(-W/2, \varphi')
     \: ,
   \\
   \\
   \displaystyle
   c_r = \frac{1}{2} \int _{-\pi/2} ^{\pi/2}
   d \varphi' \: \cos\varphi' \, f(W/2, \varphi')
   \:.
   \end{array}
\end{equation}
Boundary conditions with coefficients~\eqref{c_12} mean that the probability of electron reflection at a rough edge is independent of the scattering angle $\varphi$ and the transverse component of the electron flow $ j_y(y) = (n_0/\pi m) \int _0 ^{2\pi} d \varphi' \: \cos \varphi' \, f(y, \varphi') $ vanishes at the edges: $j_y(y=\pm W/2) =0 $ (therefore, everywhere in the sample, due to the continuity equation $\mathrm{div}\,  \mathbf{j} = j' = 0$).

The current density along the sample $j(y) \equiv j_x(y) $ is given by
 \begin{equation}
 \label{def_j}
  j(y) = \frac{n_0}{\pi m} \int _{0} ^{2 \pi }
  d \varphi' \: \sin\varphi'  \, f(y ,\varphi' )
 \:.
 \end{equation}
If an electric current flows through the sample in an external magnetic field, then, under the action of the Lorentz force, the charge-density perturbation and the Hall electric field arise. Both effects are described by the zeroth ($m=0$)  angular harmonic of the distribution function
\begin{equation}
\label{zero_harm_def}
 f^{m=0} (y) = \frac{1}{2\pi} \int _0 ^{2\pi}
d\varphi'
 \:
f(y,\varphi')
 \:.
\end{equation}

Figure~\ref{Fig1}b shows the regions of the $(y,\varphi)$ plane which correspond to the ballistic motion of electrons reflected from the right and left sample edges. In the narrow samples ($W \ll R_c$), the shape of the ''left'' and ''right'' ballistic regions is close to rectangular $[-\pi/2,\, \pi/2]\times [-W/2, \,W/2] $ and $[\pi/2,\, 3\pi/2]\times [-W/2,\, W/2] $. In wider samples ($W \sim R_c$,  $W < 2 R_c$) the boundaries of the ''left'' and ''right'' regions given by the curves $\varphi_{+} (y)$ and $ \pm \pi + \varphi_{-} (y) $, respectively (where $ \varphi_{\pm} (y) = \arcsin [1- \omega_c (W/2 \pm y ) ] $), become dependent on the $y$ coordinate (see Fig.~\ref{Fig1}b). The boundary lines coincide with the trajectories of electrons falling tangentially to the edges. Such electrons on the boundary trajectories are not described by the above boundary conditions. Consequently, the distribution function $ f(y,\varphi) $ is not defined at the points $\varphi_{+} (y)$ and $ \pm \pi + \varphi_{-} (y) $ and can therefore have a discontinuity
or another singularity at these points.

It is convenient to rewrite kinetic equation~(\ref{kin_eq}) as
\begin{equation}
\label{kin_eq_with_gamma}
 \begin{array}{c}
 \displaystyle
\Big[\cos\varphi \, \frac{\partial }{ \partial y }
  + \gamma \Big] \widetilde{f} - \sin\varphi \,E_0
 =
 \\
 \\
 \displaystyle
 = \gamma_{ee} \hat{P} [\widetilde{f} \, ]
 +
 \gamma_{d} \hat{P}_0 [\widetilde{f} \, ] +
 \omega_c \, \frac{\partial \widetilde{f}}{ \partial \varphi }
  \:,
  \end{array}
\end{equation}
where the function $\widetilde{f} (y,\varphi)= f(y,\varphi)  +  \phi(y) $ is introduced and the departure and arrival terms of the collision integral were transferred from the left to the right. Here, $\phi$ is the electrostatic potential of the Hall electric field $E_H = -\phi'$. Indeed, it follows from Eq.~(\ref{kin_eq}) that the Hall potential $\phi(y)$ plays the same role in the transport equation as the zero harmonic of the distribution function, which is proportional to the electron density and related to the Hall potential via electrostatic equations. Thus, to take into account  $\phi(y)$ and $f^{m=0}(y)$ in uniform way, it is convenient to introduce the function  $ \widetilde{f} (y,\varphi) =  f(y,\varphi)  +  \phi(y)$.

The zero harmonic  $ \widetilde{f} ^{m=0}(y)$ of the generalized distribution function ~$ \widetilde{f}$ has the form $ \widetilde{f}^{m=0}(y)=\delta \mu (y) +  \phi(y) $, where $\delta \mu(y)$ is the perturbation of the chemical potential of electrons. For stationary (and rather slow) flows, the quantities  $\delta \mu (y) $ and $  \phi(y)$ are connected by electrostatic relations~\cite{Alekseev_Alekseeva}. For the investigated case of a one-component 2D electron gas, the electrostatic potential $\phi$ is usually significantly larger than the corresponding chemical potential perturbation  $\delta \mu$~\cite{Alekseev_Alekseeva}. Therefore, the Hall electric field is calculated using the simple formula $ E_H (y) \approx   - [ \widetilde{f} ^ {m=0} ]'(y) $.

For brevity, we hereinafter omit the tilde in the function  $ \widetilde{f} $ and use simply $ f \equiv  \widetilde{f}$.


\section{General solution in the ballistic regime}

As was shown in~\cite{pohozaja_statja,a}, in the presence of weak electron--electron scattering, in almost all magnetic fields below the critical one ($B<B_c$), specifically, when $ 2 -\omega_c W  \gg \gamma W$, the electron flow is ballistic. In this case, in the main order in the scattering rate $\gamma$, such flows are described by kinetic equation~(\ref{kin_eq_with_gamma}) with an omitted arrival term~\cite{pohozaja_statja,a}:
 \begin{equation}
\label{kin_eq_B_main}
\left[\cos\varphi \,  \frac{\partial }{ \partial y }
  + \gamma \right] f
  -\sin \varphi \, E_0
 =   \omega_c \frac{\partial f}{ \partial \varphi }
 \:.
\end{equation}
The analysis of Eq.~\eqref{kin_eq_B_main} showed that the ballistic regime is divided into three subregimes~\cite{a}. The transition between them is accompanied by the evolution of the current density $j(y)$  and the Hall electric field $E_H(y)$  profiles and the change in the type of magnetic-field dependencies of the longitudinal $\varrho_{xx} (B)$ and Hall $\varrho_{xy}(B)$ resistances.

These three subregimes are as follows.

(i) $\omega_c \ll \gamma^2 W $.
The length of the maximum cyclotron orbit segment that can be adjusted in a strip $l_b^{(2)} =\sqrt{R_c W}  $ is larger than the average electron mean free path relative to bulk scattering:~$l_b^{(2)}  \gg l=1/\gamma$. Therefore, bulk scattering determines the effective ballistic trajectory length for most electrons:~$l_b^{(1)} \sim l$.

(ii) $ \gamma^2 W  \ll \omega_c  \ll  1/W $.
In this subregime, on the contrary, we have $l_b^{(2)}  \ll 1/\gamma$, while the magnetic parameter $ \omega_c W $ is small. The maximum length of the ballistic trajectory is determined by the geometry of the trajectories and therefore is equal to~$l_b^{(2)}$.

(iii)  $\omega _c \sim   1/W   $ under the condition $    2 -\omega_c W  \gg \gamma W $.
In this case, we also have $l_b^{(2)}  \ll 1/\gamma$, but the magnetic parameter $ \omega_c W \sim 1 $ is about unity, or close to the critical value $ \omega_c W  = 2$. The average ballistic trajectory
length $l_b^{(3)}$ is also determined by the geometry of electron trajectories and has the same order of magnitude as~$W$.

An analytical solution of kinetic equation~(\ref{kin_eq_B_main}), which describes all three subregimes (i)-(iii) was obtained in~\cite{pohozaja_statja,a} using the method of characteristics for ordinary differential equations. The obtained solution $f(y,\varphi)$ is a discontinuous function with continuity domains shown in Fig.~\ref{Fig1}b. For the distribution functions of electrons reflected from the left and the right sample edges, the trajectories of which lie within $-\pi + \varphi_{-} (y) < \varphi < \varphi_{+} (y)$ and  $\varphi_{+} (y) < \varphi < \pi   + \varphi_{-}(y)$, respectively, we use $f(y,\varphi)= f_{\pm}(y,\varphi) $ [see Fig.~\ref{Fig1}b]. These two components of function $f$ have the form
\begin{equation}
\label{f_gen}
   \begin{array}{c}
   \displaystyle
f_{\pm}(y,\varphi) = \frac{E_0}{ \gamma^2 +\omega_c ^2 } \, \Big[ \,
 \omega_c \cos \varphi + \gamma \, \sin \varphi   +
 \\ \\
 \displaystyle
 + e^ {  \gamma  \varphi  / \omega_c } Z_{\pm}(\sin \varphi  +\omega_c y)
\Big]
 \:,
   \end{array}
\end{equation}
where the position-independent contributions  $ \omega_c \cos \varphi $ and $\gamma  \sin \varphi$ are particular solutions of Eq.~(\ref{kin_eq_B_main}), corresponding to the Drude formulas (when the scattering rate  $\gamma$ corresponds to scattering at disorder only). The contribution  $ e^{\gamma\varphi /\omega_c} Z_{\pm}(X)$ is the general solution of kinetic equation ~(\ref{kin_eq_B_main}) without the field term $\sin \varphi E_0$. It allows to satisfy the correct boundary conditions with nonzero parameters $c_{l,r}$ given by~(\ref{c_12}).

Substituting Eq.~(\ref{f_gen}) into the boundary conditions $f(y=\mp W/2)=c_{l,r}$, we obtain the explicit form of $Z_{\pm}(X)$~\cite{a}
\begin{equation}
   \label{Z_pl}
   \begin{array}{c}
   \displaystyle
Z_{+}(X) = e^{- \gamma \, \arcsin (X + \omega_c W /2 ) / \omega_c } \, \Big[ \,\,c_l -
   \\
   \\
   \displaystyle
   - \gamma\,(X + \omega_c W /2) - \omega_c \sqrt{ 1 - (X + \omega_c W /2)^2}  \:\Big]
   \end{array}
\end{equation}
and
\begin{equation}
   \label{Z_mi}
   \begin{array}{c}
   \displaystyle
Z_{-}(X) = e^{- \gamma \,\big [\, \pi -\arcsin (X - \omega_c W /2 ) \big ] / \omega_c } \,
\Big[ \,\,c_r -
   \\
   \\
   \displaystyle
   - \gamma\,(X - \omega_c W /2) + \omega_c \sqrt{ 1 - (X - \omega_c W /2)^2 }  \:\Big]
   \:,
   \end{array}
\end{equation}
where the coefficients $c_l$ and $c_r$ are determined by the balance boundary conditions $j_y(y=\mp W/2)=0$, which take the form
\begin{equation}
   \label{syst_c12_exact}
   \left(
   \begin{array}{cc}
      I_{ll} & I_{lr} \\
      I_{rl} & I _{rr}
    \end{array}
   \right)
    \left(
    \begin{array}{c}
        c_l \\
        c_r
      \end{array}
   \right) =- \left(\begin{array}{c}
           I_l \\
           I _r
         \end{array}
   \right)
   \:.
\end{equation}
In this equation, the coefficients in the first row of the matrix are expressed in the form of the integrals
\begin{equation}
 \label{I_ll}
I_{ll} = 2 + \int \limits _{ -\pi + \varphi _ 0 } ^{ - \pi /2 }
 d\varphi \: \cos \varphi \, e^{\gamma \, (\pi + 2 \varphi )/\omega_c }
 \:
\end{equation}
and
\begin{equation}
    \label{I_lr}
  \begin{array}{c}
  \displaystyle
I_{lr} =
\int  \limits  _ {  \pi /2 } ^ { \pi + \varphi _ 0 }
d\varphi \,  \cos \varphi
  e^{\gamma \, \big[\,   \varphi - \pi
+ \arcsin ( \sin \varphi - \omega_c W  )   \big]/\omega_c }
\:,
  \end{array}
\end{equation}
while the first component on the right-hand side is
\begin{equation}
 \label{I_l}
\begin{array}{c}
\displaystyle
I_l
 = \frac{\pi \omega_c}{2 } +
\int \limits  _{  \varphi _ 0 } ^{ \pi /2 }
 d\varphi \, \cos \varphi
 \times
\\
\\
\displaystyle
 \times
  e^{\gamma  ( 2 \varphi -\pi )/\omega_c }
 ( \omega_c \cos \varphi - \gamma \sin  \varphi  )
 -
  \\
 \\
 \displaystyle
 -
 \int  \limits  _ {  - \pi /2 } ^ { \varphi _ 0 }
d\varphi \: \cos \varphi \: e^{\gamma \, \big[\,    \varphi
- \arcsin ( \sin \varphi + \omega_c W  )  \,  \big]/\omega_c }   \times
\\
 \\
 \displaystyle
\big[
 \omega_c \sqrt{1- (\sin \varphi +\omega_c W )^2 }
  +
  \gamma \, ( \sin  \varphi + \omega_c W)
  \big]
  \: .
 \end{array}
\end{equation}
The remaining coefficients in Eq.~(\ref{syst_c12_exact}), i.e. $I_{rr} $,  $I_{rl} $ and $I_r$ are related to $I_{ll}$, $ I_{lr}$ and $ I_l$ as $I_{rr} =- I_{ll}$, $I_{rl} =- I_{lr}$ and $I_r = I_l$. In Eqs.~(\ref{I_ll})-(\ref{I_l}), the definition $\varphi _0 = \arcsin (1- \omega_c W )$ is introduced.

For an arbitrary value of $\omega_cW$, the integrals (\ref{I_ll})-(\ref{I_l}) can be calculated only numerically. The explicit expressions for them and, consequently, for the quantities $c_{l,r}$ and $j(y) $, $E_H(y)$ can be obtained analytically in the following limiting cases: $\omega_c  \ll \gamma ^2W$ (see~\cite{we_6_2},\cite{a}) for subdomain (i) of the ballistic regime; $ \gamma^2 W \ll \omega_c \ll 1/W $ (see~\cite{pohozaja_statja,a} and Section~\ref{Sec4}) for subdomain (ii); and in the right-hand singular part of subdomain (iii) when $ \gamma W \ll  2-\omega_cW \ll 1 $ (see~\cite{a} and Section~\ref{Sec4}).

Subregime (i) was studied in detail in~\cite{a,we_6,we_6_2} by solving kinetic equation~(\ref{kin_eq_B_main}) using the method of successive approximations and by analyzing the exact
solution of~(\ref{f_gen}). In this subregime, the external magnetic field introduces only small corrections to the central part of the flow in the region $|W/2 - |y|| \gg \omega_c/\gamma^2$, but at the same time, it leads to the solution for electrons at the near-edge regions $|W/2 - |y|| \lesssim \omega_c/\gamma^2$, which is not described by perturbation theory.

Subregimes (ii) and (iii) were partially studied in~\cite{pohozaja_statja,a}. Below, using the general solution of~(\ref{f_gen}), we obtain new results for the flows in subregimes (ii) and (iii). Specifically, we analytically calculate the $j(y)$ and $E_H(y)$ profiles and the magnetoresistances $\varrho_{xx}(\omega_c) $ and $\varrho_{xy}(\omega_c) $.


\section{Purely ballistic transport in moderate magnetic fields}
\label{Sec4}

In the second and third ballistic subregimes, where
\begin{equation}
\label{d}
\omega_c \gg    \gamma^2 W
   \:,\;\;\;
   2- \omega_c W \gg \gamma W
   \:,
 \end{equation}
electron dynamics described by~(\ref{c_12}) and (\ref{kin_eq_B_main}) could be controlled by applied fields, scattering at the edges or by scattering in the bulk of the sample. Indeed, in classically weak fields  ($ \omega_c \ll \gamma $) included in interval~\eqref{d}, the rate of the momentum redistribution during electron--electron collisions $\gamma$, is higher than the rate of the momentum variation caused by cyclotron rotation, $\omega_c$. This is reflected, in particular, in the fact that the first term dominates in the denominator of formula~(\ref{f_gen}). However, our analysis of Eqs.~(\ref{f_gen})-(\ref{I_l}) shows that in both subdomains  $ \gamma ^2 W   \ll \omega_c   \ll  \gamma    $    and   $ \gamma    \ll \omega_c   \lesssim 1/W $ of interval~\eqref{d}, the electron distribution function in the main order in $\gamma$ is determined by scattering at the edges and the effect of applied fields, while the electron--electron collisions only lead to minor corrections.

This analysis is based on asymptotic expansion of the argument of the exponent in functions $f_{\pm}$~(\ref{f_gen})--(\ref{syst_c12_exact}) in all ranges of $\varphi$, both for the small differences ($ |\pi/2 - |\varphi| |\lesssim \omega_c W $) and for the large ones ($ | \pi/2 - |\varphi| | \gg \omega_c W $). The form of these expansions depends on the ratios $ \gamma / \omega_c $ and  $\gamma^2 W / \omega _c $. When the latter quantity is small, the exponent argument is small in the entire range of angles and the distribution function~$f_{\pm}$~(\ref{f_gen}) in the main order in $ \gamma ^ 2 W / \omega_c \ll 1  $     (and $ \omega_c W \ll 1 $ at $\gamma \ll \omega_c$) becomes independent of $\gamma $ at any angle $\varphi$.

The negligible role of bulk scattering in the main order in $\gamma$ in both ranges $\gamma^2 W \ll \omega_c \ll \gamma $ and $\gamma \ll \omega_c \ll 1/W$ can be qualitatively explained as follows. As we noted in the previous section, at $ \omega_c \gg \gamma ^2 W $, the maximum size of the ballistic trajectory is limited not by the length $l=1/\gamma$, related to bulk scattering, but by the maximum length $ l_b^{(2)}  =  \sqrt{ R_cW} $ of the cyclotron orbit segment that can be fit to the sample (see Fig.~\ref{Fig1}a). Consequently, in the both ranges of $\omega_c$, the reflection of any electron from an edge is most likely followed by its subsequent reflection from the same or opposite edge than the bulk scattering. Such dynamics of electrons determines the current density along the $x$ axis and the Hall electric field, which is obtained from the requirement for the absence of current and average acceleration along the $y$ axis. Therefore, the current density and the Hall field in the main order in the parameter $ l_b^{(2)} / l \ll 1$  (and $W/l_b^{(2)} \ll 1 $ at $\omega_c \ll \gamma$) are described by the purely ballistic formulas which do not account for the scattering in the bulk.

Thus, over the entire range~(\ref{d}), Eq.~(\ref{f_gen})  at  $\gamma = 0$ yields the desired distribution function in the main order in the bulk-scattering rate $\gamma $ for all $y$ and $ \varphi$:
\begin{equation}
\label{f_bez_gamma}
 \begin{array}{c}
 \displaystyle
f_{\pm} (y, \varphi )= \widetilde{c} _{l,r}
 + \frac{E_0}{\omega_c }
 \Big\{ \,
 \cos \varphi
 \,
 \mp
 \\\\
 \displaystyle
 \mp
 \sqrt{
  1-\Big[\sin \varphi + \omega_c \, \Big(y \pm
  \frac{ W}{2}\Big ) \, \Big]^2
 }
 \; \Big\},
 \end{array}
\end{equation}
where  $\widetilde{c} _{l,r}  = E _0  c _ {l,r} / \omega_c^2 $. We note that, at $\gamma ^2 W \ll  \omega_c \ll \gamma$ taking the limit $\gamma \to 0 $ in function~(\ref{f_gen}) is impossible and the coincidence of the final form of the function $f_{\pm}$ at $\gamma^2W\ll \omega_c \ll \gamma$ with~(\ref{f_bez_gamma}) in the main order in the parameter $\omega_c W \ll 1$ is the result of analysis of the first two terms of the asymptotics of general expression~(\ref{f_gen}) in the parameter $\gamma^2 W / \omega_c  \ll 1 $ under the condition  $\omega_c W \ll 1 $.

Any solution of Eq.~(\ref{kin_eq_with_gamma}) at $\gamma =0$ can be found accurate to a constant, which corresponds to the absence of relaxation of electron-density perturbations. Consequently, the system of algebraic equations for coefficients $c_{l,r}$~(\ref{syst_c12_exact}) is degenerate. Imposing the symmetry condition $c _ {l}+ c _ {r} =0$, we find from system~(\ref{syst_c12_exact})
\begin{equation}
\label{c_bez_gamma}
\begin{array}{c}
\displaystyle
  \widetilde{c} _{l,r}
  =\mp
  \frac{E_0}{\omega_c}
  \frac{U - V}{2\,(2-\omega_c W)}
  \:,
   \end{array}
\end{equation}
where $   U=   \arccos (1-\omega_cW )$ and $  V =   (1-\omega_cW )\sqrt{\omega_cW} \sqrt{2-\omega_c W}$. We note that the form of solution~(\ref{f_bez_gamma}) was obtained recently in~\cite{pohozaja_statja,a}, but its range of applicability was not fully analyzed there.

As can be seen from Eq.~(\ref{f_bez_gamma}), for intermediate magnetic fields at $\omega_c \sim 1/W$ (the left-hand side of subregime (iii)), the current density $j(y)$ is estimated as
\begin{equation}
  \label{j_full_ball}
  j(y)\sim j_0
\end{equation}
where $j_0 = n_0E_0W/m$ is the characteristic density of the purely ballistic current. Equation~(\ref{j_full_ball}) follows from the fact that the typical ballistic trajectories, which give the main contribution to the current $j$ at $\omega_c \sim 1/W$, have the length of the order of $W$. In the same fields,  $\omega_c \sim 1/W$, the distribution function~(\ref{f_bez_gamma}) also allows us to estimate the Hall field
\begin{equation}
    \label{E_H_full_ball}
    E_H(y)\sim E_0
    \:.
\end{equation}

\begin{figure}[t!]
\includegraphics[width=0.95\linewidth]{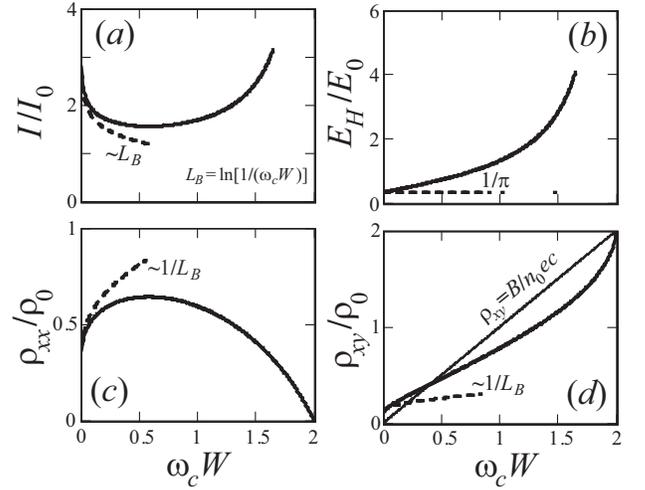}
\caption{(a) Total current, (b) Hall electric field, and (c) longitudinal and (d) Hall resistance for a sample without bulk scattering ($\gamma = 0$) as functions of the magnetic field in the second ($0<\omega_c W \ll 1$) and the third ($1 \lesssim \omega_c W <2 $) ballistic subregimes. The resistances are given in units of $\varrho_0  = E_0 / j_0 = m/(n_0W)$.}
\label{Fig2}
\end{figure}

In Fig.~\ref{Fig2}, we present the calculated dependencies of the total current $ I = \int_{-W/2} ^{W/2} dy \, j(y) $, the averaged Hall electric field $ E_H =\int_{-W/2} ^{W/2} (dy/W ) \, E_H(y) $, and the resistances $ \varrho_{xx} = E_0/I$ and $\varrho_{xy} = E_H/I$ on the parameter $\omega_c W$ over the entire ballistic region $0<\omega_c<2/W$ with the absence of bulk scattering ($\gamma=0$). It can be seen in Fig.~\ref{Fig2} that estimates (\ref{j_full_ball}) and (\ref{E_H_full_ball}) of the current  $I$ and the Hall field $E_H$ are, in fact, valid in the middle part of the region, where    $\omega_c \sim 1/W$. However, the current  $I(\omega_c)$ diverges at both edges of the interval and the Hall field $E_H(\omega_c)$ diverges at the right-hand edge  ($\omega_c W \to 2 $), remaining finite at the left-hand edge  $\omega_c \to 0 $.

The existence of a finite value of the Hall field  $ E_H \to E_0 / \pi $ in the weak-field limit $\omega_c \to 0$ is a nontrivial manifestation of the ballistic nature of electron motion. A similar behavior of the  $\varrho_{xy}(B)$ dependence was obtained earlier in experiments and the numerical simulation~\cite{Beenakker_Houten_obz} of 2D electrons in samples with four contacts and was called ballistic anomalies of magnetoresistance. In terms of review~\cite{Beenakker_Houten_obz}, the $\varrho_{xy}(B)$ in Fig.\ref{Fig2}d in the range $\omega_c W \ll 1 $ is the ''last Hall plateau''.

Next, we obtain analytical expressions for the flow characteristics within the $\omega_c \to 0 $ and $\omega_c W \to 2 $, i.e., in subregime (ii) and the right-hand singular part of subregime (iii).

According to distribution function~(\ref{f_bez_gamma}), in the weak magnetic-field limit $ \omega_cW \ll 1 $, the main contribution to the transport characteristics is made by electrons moving along the edges of the sample with angles $\varphi \approx \pm \pi/2$. The asymptotic form of Eq.~(\ref{f_bez_gamma}) at angles $ \sqrt{\omega_cW } \ll |\pi/2 - |\varphi| | \ll 1$ is
 \begin{equation}
    \label{c_bez_gamma__next}
  \begin{array}{c}
                     \displaystyle
                          f_{\pm} (y, \varphi ) =  \widetilde{c} _{l,r} +
                          E_0\, \Big[ (y\pm W/2 ) \, \frac{\sin \varphi }{ \cos \varphi} +
                          \\
                          \\
                          \displaystyle
                          + \omega_c\frac{ (y\pm W/2 )^2 }{2} \Big( \,
                          \frac{1 }{ \cos \varphi}
                          +
                          \frac{\sin ^2 \varphi }{ \cos ^3\varphi} \, \Big)
                          \Big]
                         \:,
                         \end{array}
                        \end{equation}
where
 \begin{equation}
  \widetilde{c} _{l,r} = \pm \frac{ \sqrt{2} } {3 } \, E_0 \, \sqrt{ \omega_c  \, W ^{3}} \:.
 \end{equation}
At the angles $ |\pi/2 - |\varphi| |  \lesssim \sqrt{\omega_cW } $, it is necessary to use the complete expression~(\ref{f_bez_gamma}), in which $\sin \varphi  \approx \pm [ \,  1-( |\varphi| -\pi /2 )^2/2 \, ] $ and $ \cos\varphi  \approx \pm  | |\varphi| -\pi /2 |  $.

\begin{figure}[t!]
 \includegraphics[width=0.95\linewidth]{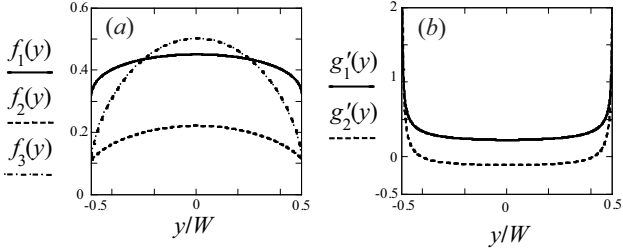}
 \caption{Functions determining (a) the current density and (b) Hall electric field profiles according to Eqs.~(\ref{j_0})-(\ref{g2}) in the second ballistic sub-regime $ ( \gamma W )^2  \ll \omega_c W  \ll  1 $, obtained in the absence of interparticle scattering.}
\label{Fig3}
\end{figure}

The current density~(\ref{def_j}), corresponding to distribution function~(\ref{c_bez_gamma__next}), consists of the main part independent of the $y$ coordinate and a small correction to the logarithmic parameter $L_B=\ln[1/(\omega_cW)] \gg 1 $ dependent on $y$
  \begin{equation}
   \label{j_0}
   \begin{array}{c}
   \displaystyle
   j(y)  =j _{B}  + \Delta j(y)\:,
 \quad
   \frac{ j _B   } {j_0}
     = \frac{1}{\pi} \,
   \ln\Big( \frac{1}{\omega_c W }\Big)
   \:,
   \\
   \\
   \displaystyle
  \frac{ \Delta j(y)} {j_0} =
  f_1(y)+f_2(y)+f_3(y) \:.
    \end{array}
  \end{equation}
The  $f_1(y)$, $f_2(y)$ and $f_3(y)$  functions were calculated analytically. They have similar profiles with a divergent coordinate derivative at the sample edges $y =\pm W/2$:
  \begin{equation}
   \label{Dj1}
   f_1(y)=\frac{1}{\pi} \Big( \,
  \sqrt{\frac{1}{2} + \frac{y}{W}} +\sqrt{\frac{1}{2} - \frac{y}{W}}
   \: \Big)
  \:,
  \end{equation}
  \begin{equation}
   \label{Dj2}
      \begin{array}{c}
   \displaystyle
  f_2(y)=
    \frac{1}{\pi} \Big\{ \,   \Big(\frac{1}{2}+\frac{y}{W}\Big) \, \ln \Big[ \frac{\sqrt{2 }}{
   \sqrt{ \frac{1}{2}+\frac{y}{W}}
   }\Big]+
  \\
  \\
   \displaystyle
   +
   \Big(\frac{1}{2}-\frac{y}{W}\Big) \, \ln \Big[ \frac{\sqrt{2}}{\sqrt{
   \frac{1}{2} - \frac{y}{W}}
   } \Big]  \, \Big\}
   \:,
   \end{array}
  \end{equation}
   and
  \begin{equation}
   \label{Dj3}
   \begin{array}{c}
   \displaystyle
  f_3(y)=
   \frac{1}{\pi} \Big\{ \,
   \Big(\frac{1}{2}+\frac{y}{W}\Big) \, \ln \Big[ \frac{\sqrt{2}}{
  1-\sqrt{
   \frac{1}{2}-\frac{y}{W}}
   }\Big]+
  \\
  \\
  \displaystyle
   +
   \Big(\frac{1}{2}-\frac{y}{W}\Big) \, \ln \Big[ \frac{\sqrt{2}}{
   1-\sqrt{
   \frac{1}{2} + \frac{y}{W}}
   }\Big]
   \, \Big\}
    \:.
    \end{array}
  \end{equation}
The similarity of the profiles can be seen in Fig.~\ref{Fig3}a, which shows the $f_1(y)$, $f_2(y)$ and $f_3(y)$ dependencies.

The Hall-field potential~(\ref{zero_harm_def}) calculated from the distribution function~(\ref{c_bez_gamma__next}) takes the form
    \begin{equation}
     \label{phi}
    \phi(y) = E_0 W [\,g_1(y) + g_2(y) \,] \:,
    \end{equation}
where
   \begin{equation}
    \label{g1}
     g_1(y)=
    \frac{1}{2\pi}
     \Big(\,
    \sqrt{\frac{1}{2} + \frac{y}{W}} -\sqrt{\frac{1}{2} - \frac{y}{W}}
     \: \Big)
    \end{equation}
 and
     \begin{equation}
     \label{g2}
     \begin{array}{c}
     \displaystyle
     g_2(y)=
     \frac{1}{2\pi}   \Big\{
     \,
     \Big(\frac{1}{2}+\frac{y}{W}\Big) \, \ln \Big[\, \frac{
     1-\sqrt{
    \frac{1}{2}-\frac{y}{W}}
      }{ \sqrt{\frac{1}{2}+\frac{y}{W}} }\, \Big]+
    \\
    \\
    \displaystyle
   -
  \Big(\frac{1}{2}-\frac{y}{W}\Big) \, \ln \Big[\, \frac{
   1-\sqrt{
   \frac{1}{2}+\frac{y}{W}}
   }{ \sqrt{\frac{1}{2}-\frac{y}{W}} }\, \Big]
    \, \Big\}
    \:.
    \end{array}
   \end{equation}
These formulas describe the Hall field in the main order in the parameter  $\omega_cW \ll 1$. Both functions  $g_1(y)$ and $g_2(y)$ have derivatives divergent near the sample edges, which leads to divergence of the Hall field $E_H (y) =- \phi' (y)$ (see Fig.~\ref{Fig3}b).

\begin{figure}[t!]
 \includegraphics[width=0.95\linewidth]{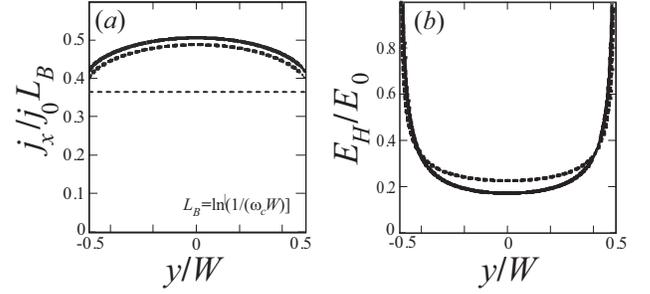}
 \caption{(a) Current density and (b) Hall field at $\gamma = 0$ in the second ballistic subregime $\omega_cW  \ll 1 $ (specifically, $ \omega_c W   = 10^{-3}$). Solid lines are the numerical results obtained by integrating distribution function~(\ref{f_bez_gamma}). Analytical results of Eqs.~(\ref{j_0}) and (\ref{phi}): (a) thick dashed line, (b) the analytical calculation by Eq.~(\ref{phi}) coincides with the thick solid line. The thin dashed curve in (a) shows the main uniform contribution to  $ j(y) \approx  j_B $. The difference between the solid and thick dashed lines for the current density arises from the corrections in the parameter  $1/L_B$, which are about unity and not taken into account in formula~(\ref{j_0}). The dashed curve in (b) corresponds to the contribution proportional to the derivative $g_1'(y)$ in~(\ref{phi}), which is the main contribution to the Hall field $E_H(y) $, since the derivative $g_2'(y)$ is relatively small as compared with $g_1'(y)$.  }
\label{Fig4}
\end{figure}

In Fig.~\ref{Fig4}, we compare the numerically calculated $j(y) $ and $E_H (y)$ profiles using exact distribution function~(\ref{f_bez_gamma}) and analytical expressions~(\ref{j_0}) and~(\ref{phi}). It can be seen that analytical curves~(\ref{j_0}) and~(\ref{phi}) correctly reproduce the results of the numerical calculation using~(\ref{f_bez_gamma}) for the current and the Hall field in the limit  $\omega_cW \ll 1$. The current density $j(y)$ calculated accurately contains a correction of the order of unity independent of  $y$, which is not taken into account in formulas~(\ref{j_0})-(\ref{Dj3}); the numerically obtained Hall field is accurately reproduced by Eqs.~(\ref{phi})-(\ref{g2}).

Using Eqs.~(\ref{j_0})-(\ref{g2}) in the main order in $\omega_c W $, we obtain the following results for the total current $I$ and the Hall voltage $U_H = \phi(W/2) - \phi(- W/2) $:
     \begin{equation}
     \label{I__U_H}
     I=
      j_0 W \frac{L_B}{\pi}
\:,\quad
     U_H =
   \frac{1}{\pi} E_0 W
      \:.
   \end{equation}
Consequently, the longitudinal and Hall resistances are
\begin{equation}
 \label{resist_ii}
\varrho_{xx} (B) = \varrho_ 0 \frac{1}{L_B}
\:,\quad
\varrho_{xy} (B) = \varrho_ 0\frac{1}{\pi \, L_B} \:,
   \end{equation}
where $ \varrho_ 0  = E_0 /j_0 = m / (n_0 W) $. It can be seen that both these quantities have a weak singularity due to the presence of the factor  $ 1/L_B(B) \sim 1/\ln(1/B)$.

\begin{figure}[t!]
   \includegraphics[width=0.95\linewidth]{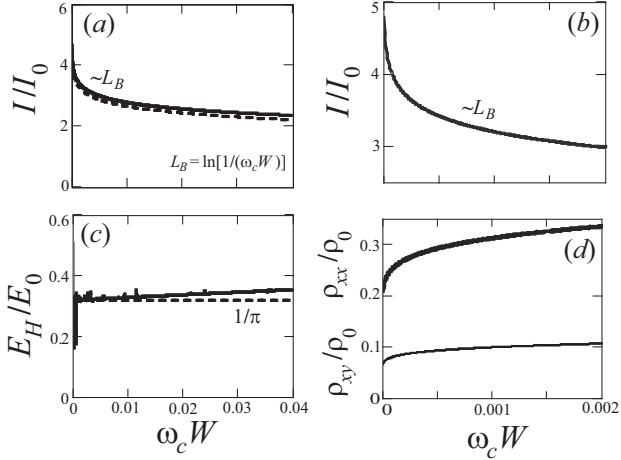}
   \caption{(a, b) Total current, (c) Hall electric field, and (d) longitudinal and Hall resistances calculated at  $\gamma = 0$ as functions of the
magnetic field in the second ballistic sub-regime $0<\omega_cW\ll1$. Solid curves are the numerical results calculated using distribution function~(\ref{f_bez_gamma}) and dashed lines are calculated using analytical expressions (\ref{I__U_H}) and (\ref{resist_ii}). (d) The thin curve corresponds to the Hall resistance and the thick curve, to the longitudinal resistance.}
\label{Fig5}
\end{figure}

Figure~\ref{Fig5} shows the dependencies of the total current, the average Hall field, and the corresponding resistances in the second ballistic sub-regime $\omega_c W \ll 1$ on the parameter $\omega_cW$. We compare the results of numerical calculation of all the quantities using distribution function~(\ref{f_bez_gamma}) with the analytical results of~(\ref{I__U_H}) and~(\ref{resist_ii}). It can be seen that the latter describe the numerical calculation well. The singularities in the resistances $\varrho_{xx}$ and  $\varrho_{xy}$ at $B \to 0 $ presented in Fig.~\ref{Fig5} are very weak.

Near the transition, at $ \gamma W  \ll  2 - \omega_cW  \ll 1 $ (the right-hand side of the third ballistic subregime), coefficients $\widetilde{c} _{l,r}$~(\ref{c_bez_gamma}) rapidly diverge. In the main order in the parameter $ u = 2- \omega_c W   \ll 1 $, they take the form $ \widetilde{c} _{l,r}  =   \pm (\pi E_0)/[2 \, \omega_c \, (2-\omega_cW)] $ and become larger than the other terms of Eq.~(\ref{f_bez_gamma}). Then, the main part of the distribution function is~\cite{a}
 \begin{equation}
 \label{f_at_2}
  f_{\pm} (y,\varphi)
   \approx \pm \frac{\pi E_0}{2  \omega_c  u}
     \:,
   \end{equation}
The terms omitted in this formula are of the order of $E_0/\omega_c$.

Distribution function~(\ref{f_at_2}) describes the disbalance between the densities of excess electrons reflected from the left and right edges, which is needed to compensate the nonequilibrium current $ j_y $ induced by the direct action of the fields  $\mathbf{E}_0$ and $\mathbf{B}$ on electrons [the first two terms in formula~(\ref{f_gen})].

\begin{figure}[t!]
   \includegraphics[width=0.95\linewidth]{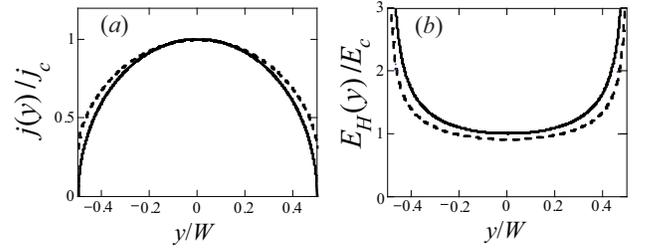}
   \caption{(a) Current density and (b) Hall electric field in the sample without bulk scattering ($\gamma = 0 $) for the right-hand side of the third ballistic sub-regime $0 < 2-\omega _cW  \ll 1 $.  The solid curves correspond to magnetic fields very close to the transition point $u \to 0 $ and dashed curves, to the parameter  $u=0.2$. The current density is normalized to $j_c =2E_0 / (\omega_c u)  $ and the Hall field is normalized to  $E_c = E_0/u$.}
\label{Fig6}
\end{figure}

Calculation by formulas~(\ref{def_j}) and~(\ref{zero_harm_def}) with distribution function~(\ref{f_at_2}) yields the current density
 \begin{equation}
    \label{j_nea_c}
   \begin{array}{c}
   \displaystyle
j(y ) =
 \frac{ E_0 }{\omega_c u}
 \Big[ \,
\sqrt {1-\Big(\omega_c y -\frac{u}{2}\Big)^2}
 +
  \sqrt {1-\Big(\omega_c y + \frac{u}{2}\Big)^2}
 \: \Big]
\end{array}
\end{equation}
and the electrostatic potential of the Hall field
 \begin{equation}
 \label{EH_nea_c}
   \begin{array}{c}
   \displaystyle
\phi_H(y ) =
  \frac{E_0}{2 \omega_c u } \,
 \Big[\,
\arcsin\Big(\omega_c y -\frac{u}{2}\Big)
  +
   \arcsin\Big(\omega_c y +\frac{u}{2}\Big)
 \,\Big]
 \:.
\end{array}
\end{equation}
Both quantities~\eqref{j_nea_c} and~\eqref{EH_nea_c}, as well as distribution function~(\ref{f_at_2}), diverge as $\sim 1/u$ in magnetic fields close to the transition point $ \gamma W  \ll  2 - \omega_cW  \ll 1 $. The current density profile is a distorted semicircle with very large derivatives at the sample edges $y=\pm W/2$ and the Hall potential profile has the form of a distorted arcsine. These quantities for the sample without bulk scattering ($\gamma = 0 $) are shown in Fig.~\ref{Fig6} for the transition point $u=0$ ($B=B_c$) and just below it $u\ll 1$ ($B_c-B \ll B_c $).

Using~(\ref{j_nea_c}) and~(\ref{EH_nea_c}), we obtain the following expressions for the total current and Hall voltage in the main and subsequent orders in the parameter $\sqrt{u}$
 \begin{equation}
I=   \frac{\pi j_0 W }{ 4  u }
\: , \quad
U_H =    \frac{\pi E_0 W }{ 2 u }
\, \Big(1-
  \frac{
  \sqrt{2u}
  }{\pi}
 \Big)
 \:.
\end{equation}
The corresponding resistances are
\begin{equation}
\label{asympt}
\varrho_{xx}
= \frac{4 u}{\pi} \varrho_0
\: , \quad
 \varrho_{xy} = 2 \Big(1- \frac{\sqrt{2u}}{\pi}\Big) \,  \varrho_0 \:.
 \end{equation}
The longitudinal resistance $\varrho_{xx}$ as a function of $u $ decreases linearly to zero at $u \to 0 $, while the Hall resistance $\varrho_{xy}$ shows the root divergence in this limit. This behavior of the resistances was also obtained via the direct numerical calculation using the distribution function~(\ref{f_bez_gamma}) [see Fig.~\ref{Fig2}].

\section{Discussion of the results}

The comparison of the calculated current and Hall field in relatively low ($ \gamma^2 W^2  \ll  \omega_cW  \ll  1 $), relatively large near-critical ($ \gamma W  \ll  2-\omega_cW  \ll 1  $), and intermediate ($\omega_cW \sim  1 $) magnetic fields shows that, with an increase of magnetic field, the current density profile becomes increasingly convex and, for the Hall field profiles, the width of the diverging features at the sample edges increases (the profiles for the first and second cases are shown in Figs.~\ref{Fig4} and ~\ref{Fig6} and, for the third case, the shapes of the $j(y)$ and $E_H(y)$  curves are intermediate between the corresponding curves in Figs.~\ref{Fig4} and~\ref{Fig6} and therefore are not shown). For the current density, the ratio between the uniform $j(y)$ part and the nonuniform one decreases with an increase in $\omega_c$ in the range of $ \gamma^2 W^2   \ll \omega_cW  \ll  1 $ as a logarithm $\sim \ln[1/(\omega_c W)]$ [see formulas~(\ref{j_0})]. The last formula, as well as numerical calculation of the current using distribution function~(\ref{f_bez_gamma}), show that the uniform and nonuniform $j(y)$ portions are of the same order of magnitude as  $\omega_c W  \sim 1 $.

We note the nontrivial features of the obtained magnetic-field dependencies of the longitudinal and Hall resistances $\varrho_{xx}$ and $\varrho_{xy}$.

First, both these functions are nonanalytical in the limit $\omega_c W \ll 1 $: $ \varrho_{xx/xy} (\omega_c) \sim 1/\ln[1/(\omega_c W ) ]$. A strongly singular behavior of this type is not encountered in the ohmic- and hydrodynamic transport modes and is characteristic for ballistic transport, when the main contribution to the current is given by a group of trajectories of selected geometry. In the investigated system, in the limit $\omega_c W \ll 1$,  such a group of trajectories are segments of cyclotron circles with velocity angle close to the sample direction: $|\varphi| \approx \pi /2  $.

Second, it is noteworthy that, in the lower vicinity of the transition, the Hall resistance $\varrho_{xy}$ takes exactly the nominal ballistic value $\varrho_0$ and the longitudinal resistance $\varrho_{xx}$ tends to zero linearly with respect to the difference $B_c-B$ (for the case $\gamma \to  0$). The vanishing of $\varrho_{xx}$ at $B \to B_c$ is consistent with the fact that 2D electrons in a perpendicular magnetic field in a strip with a width equal to or greater than the critical value ($W \geq 2R_c$) in the complete absence of disorder and electron--electron interaction ($\gamma = 0 $) do not have a finite
resistance. In such a system, at $W \geq 2R_c$, the applied external field $\mathbf{E}_0 || \mathbf{e}_x $, at a certain instant of time $t=0$, leads to unlimited growth of all the quantities in the system: $j(y,t) = j_x(y,t) $ along the external field with time, as well as the current along the normal to the sample edge $j_y(y,t)$ and, consequently, the Hall field $E_H(y,t)$.

The calculated dependencies of the resistances $\varrho_{xx}(B)$ and $\varrho_{xy} (B) $ in the range of  $\gamma ^2 W ^2   \ll  \omega_cW  <2  $ [Figs.~\ref{Fig2}c,~\ref{Fig2}d; formulas~(\ref{resist_ii}) and~(\ref{asympt})] are in good agreement with the behavior of the resistances obtained by the numerical solution of kinetic equation~(\ref{kin_eq}) for the case of weak interparticle scattering $\gamma W \ll 1 $ (see Figs.~1a and~2a in~\cite{Scaffidi}).

The coincidence of the features of the obtained theoretical field dependence of the longitudinal resistance $\varrho_{xx}(B)$ (the maximum in the region of $\omega _c  \approx 1/ W $ and much smaller values at $\omega _c W\ll 1$  and  $\omega _c W \approx 2$)  with the features of the resistance observed at low temperatures in high-quality long samples of graphene and GaAs quantum wells (see Fig.~1 in~\cite{rrecentnest}, Fig.~S21 in the appendix to~\cite{rrecentnest2}, and Fig.~2 in~\cite{Gusev_2_Hall}) apparently indicates that our theory and the numerical theory from~\cite{Scaffidi} describe the experiments reported in~\cite{rrecentnest,rrecentnest2,Gusev_2_Hall}.


\section{Conclusions}

The ballistic flow of two-dimensional electrons in a magnetic field in long samples with rough edges was studied. It was shown that, in a wide range of magnetic fields up to the critical field of the transition to the hydrodynamic regime, the flow is mainly determined by the scattering of electrons at the strip edges and their acceleration by magnetic and electric fields. The current density and Hall electric field distributions over the sample cross section were calculated and the longitudinal and Hall resistances as functions of the magnetic field were determined. The obtained dependence of the longitudinal resistance apparently agrees with the dependencies observed experimentally in~\cite{rrecentnest,rrecentnest2,Gusev_2_Hall} in pure samples of graphene and GaAs quantum wells in magnetic fields below the transition between the ballistic and hydrodynamic transport regimes. It seems important to compare in detail the obtained ballistic profiles of the current density and Hall field with recent measurements of these quantities in graphene samples from~\cite{rrecentnest,rrecentnest2}.


\section*{Acknowledgements}

We are grateful to A.I. Chugunov for useful discussions.


\section*{Funding}

The study was supported by the Russian Foundation for Basic Research, project no. 19-02-00999.


\section*{Conflict of interest}

The authors declare that they have no conflict of interest.

\bibliography{bib_hydro}

\end{document}